\documentclass[a4paper]{jpconf}
\usepackage{graphicx}
\begin{document}
\title{Highlights of D\O\ QCD related analyses}

\author{Lars Sonnenschein {\hspace*{1ex} \rm on behalf of the D\O\ collaboration \\ (submitted to the proceedings of the Photon 2011 conference, Spa)}}

\address{RWTH Aachen University, III. Phys. Inst. A, 52056 Aachen, Germany}

\ead{Lars.Sonnenschein@cern.ch}

\begin{abstract}
D\O\ provides a wealth of measurements conceived for probing perturbative and non-perturbative aspects
of QCD, giving an accurate experimental account for Standard Model production processes including 
jets, leptons and photons and improving the sensitivity and the understanding in searches for new physics.
Among the most important subjects are inclusive jet production, vector boson plus jet 
production, direct photon production and measurements in Minimum Bias events and Double Parton 
Scattering which are discussed here and compared to theory predictions.

\end{abstract}

\section{Introduction}

Proton anti-proton collisions at a centre of mass energy of $\sqrt{s}=1.96$~TeV are taking place at the
the Tevatron accelerator, Fermilab. A peak luminosity of $4.3\cdot 10^{32}$~cm$^{-2}$s$^{-1}$ has been 
reached and over 11.6 fb$^{-1}$ integrated luminosity are already delivered, 
of which 10.5 fb$^{-1}$ have been recorded by the D\O\ experiment. 
The presented analyses are based on integrated luminosities ranging between 0.7 and 4.2 fb$^{-1}$.

The D\O\ detector \cite{d0det} has broad particle identification capabilities. 
Jets, electrons and photons are detected by the calorimeter with fine granularity
(in pseudorapidity and azimuthal angle of $\Delta\eta \times \Delta\phi \sim 0.1 \times 0.1$) 
and good energy resolution. Charged particles
are reconstructed by means of the central tracking system and muons are detected in the muon 
spectrometer as outermost detector component. The data taking efficiency is 
$\geq$
90\% (in Run~2b it amounts to about 92\%, on average). 

Jet measurements are well understood by Quantum Chromodynamics (QCD) calculations.
They can be exploited to set constraints on the non-perturbative Parton Distribution Functions (PDF's).
In the kinematic plane of proton momentum fraction $x$ and virtuality scale $Q^2$, regions accessible
to fixed target experiments, HERA, the Tevatron and LHC are complementary. Only Tevatron inclusive jet
data provide significant constraints at high $x$ and $Q^2$.
Jet measurements are also important within searches for new phenomena. Heavy particles can decay into
quarks which in turn fragment into jets. Extra dimensions and quark compositeness are examples of such new kinds of phenomena. 
  
The production of dijets at the Tevatron is dominated 
by the gluon gluon fusion process at low transverse jet momenta. 
Toward larger transverse jet momenta the fraction of quark anti-quark annihilation dominates, 
while the fraction of quark gluon fusion tends to decrease marginally.
 
Among all past and present hadron colliders the Tevatron inclusive jet production data 
dominates together with published LHC results the high jet transverse momentum reach.
At the same time, the Tevatron has by far the best sensitivity at large (anti-)proton momentum 
fractions $x$.  

While the data are corrected for detector effects,
the predictions take into account non-perturbative contributions due to the 
underlying event and hadronisation. Finally, the comparison to data is then established at the 
hadronic final state as defined in \cite{buttar08}.

\noindent
The convention to set the speed of light $c\equiv 1$ is adopted throughout this article.

The following section covers inclusive jet production measurements. Section \ref{vec+jets} shows two 
examples of vector boson plus jet production measurements. Section \ref{directPhotons} encompasses
the most relevant direct photon production measurements of D\O. Important non-perturbative measurements
in the regime of minimum bias events and double parton scattering are given in section \ref{MB+DPS}.

\section{Inclusive jet production}
 
The measurement of the cross section of inclusive jet production in hadron collisions provides 
stringent tests of QCD. Large jet transverse momenta with respect to the beam axis
($p_T$) ensure small contributions from long-distance processes and the jet production can be
calculated in perturbative QCD (pQCD). The inclusive jet cross section in $p\bar{p}$ collisions
provides one of the most direct probes of physics at small distances. In particular, it is directly 
sensitive to the strong coupling constant ($\alpha_s$) and the PDF's of the proton. 
Furthermore it can be exploited to set constraints on the internal structure of quarks
\cite{Peskin83}.

\subsection{Inclusive jet production}
The double differential inclusive jet production cross section is 
measured~\cite{d0_incl_Jets_prl101_062001_2008} as a function of the leading jet transverse momentum 
and the rapidity based on an integrated luminosity of 0.7 fb$^{-1}$.
The D\O\ Run~II iterative seed-based cone jet algorithm including mid-points~\cite{run2cone} with cone radius
$R=\sqrt{(\Delta y)^2 + (\Delta\phi)^2}=0.7$ in rapidity $y$ and azimuthal angle $\phi$ space is used 
to cluster energies deposited in calorimeter towers as well as for the pQCD calculations on the 
partonic final state.
The jet transverse momentum is corrected for the energy response of the calorimeter, 
energy showering in and out of the jet cone and for additional energy from event pile-up and 
multiple parton interactions.
Events with jets above a transverse momentum threshold of 50~GeV in the rapidity range of $|y|<2.4$
are selected. The cross section is measured in six rapidity bins as a function of jet $p_T$, reaching
transverse momenta above 600~GeV.
The cross section is corrected for muons and neutrinos, not reconstructed within jets.
Corrections for jet migration between bins in $p_T$ and $y$ due to finite resolution in energy and position are determined in an unfolding procedure, based on the experimental $p_T$ and $y$ resolutions.
The bin sizes are chosen to minimise migration corrections due to the experimental resolution.
The dominant systematic uncertainty comes from the Jet Energy Scale (JES).
The steeply falling jet transverse momentum spectrum causes shifts in energy which translate into a 
large uncertainty on the cross section. The typical JES uncertainty is 1 - 2\% which results in a 
total uncertainty on the cross section of 15 - 30\%. 
Perturbative QCD predictions to next-to-leading order (NLO) in $\alpha_s$, computed using the {\sc\scriptsize FAST}NLO program \cite{fastNLO}, based on NLOJET++ \cite{nlojet} are compared to data making use of PDF's
from CTEQ6.5M \cite{cteq6.5}\cite{cteq6.5b}\cite{cteq6.5c}.
The theoretical uncertainty due to variation of the renormalisation ($\mu_R$) and factorisation 
($\mu_F$) scales by a factor of two up and down are of order 10\%. The scale itself is chosen to 
be the individual jet $p_T$. In all $y$ regions the predictions agree well with the data. 
The data tends to be lower than the central prediction making use of CTEQ PDF's, in particular at 
very large transverse jet momentum.

\subsection{Inclusive dijet production invariant mass cross section}
The inclusive dijet production cross section is measured~\cite{d0_dijetInvmass_plb693_531_2010} 
as a function of the dijet mass based on an integrated luminosity of 0.7 fb$^{-1}$.
This measurement is sensitive to quark compositeness, extra spatial dimensions 
and undiscovered heavy particles decaying into two quarks. The invariant mass distribution 
is particularly sensitive to the PDF of gluons at high proton momentum fraction $x$
where the gluon distribution is weakly constrained.
The two leading jets (cone jet algorithm~\cite{run2cone} with radius $R=0.7$) are required to exceed a 
transverse momentum threshold of 40~GeV. The cross 
section is measured in different regions of maximal rapidity, defined as $|y|_{\max}=\max(|y_1|,|y_2|)$.
The six equidistant adjacent regions are given by:
$0<|y|_{\max}<0.4$, $0.4<|y|_{\max}<0.8$, $0.8<|y|_{\max}<1.2$, $1.2<|y|_{\max}<1.6$, 
$1.6<|y|_{\max}<2.0$ and $2.0<|y|_{\max}<2.4$. 
Calorimeter shower shapes are used to remove remaining background due to electrons, photons and 
detector noise giving rise to jets. The invariant mass bins are chosen to have a size of about 
twice the mass resolution and to respect an efficiency and purity of about 50\%. The jet $p_T$ 
resolution is measured in events with exactly two jets and has been found to be approximately 13\%
(7\%) at $p_T\simeq 50~(400)$~GeV in the Central Calorimeter (CC) and the Endcap Calorimeter (EC) and 
16\% (11\%) at $p_T\simeq 50~(400)$~GeV in the Inter Cryostat Region (ICR) between the two calorimeter 
compartments CC and EC.
The total experimental corrections to the data are less than $\pm2\%$.
The systematic uncertainties on the cross section are dominated by the uncertainties in the jet 
energy calibration which range from 6 - 22\% in the EC to 15 - 45\% in the CC.
The second largest systematic uncertainty is given by the $p_T$ resolution uncertainty, ranging between 2 and 10\% in all regions. The luminosity determination has an uncertainty of 6.1\% which is completely correlated across all bins.
The data are compared to the NLO prediction computed using \cite{fastNLO}, 
based on NLOJET++ \cite{nlojet}\cite{nlojet02} for MSTW2008NLO PDF's \cite{mstw2008} 
with $\alpha_s(m_Z)=0.120$. In addition, a comparison to theoretical predictions using
CTEQ6.6 PDF's \cite{cteq6.6} has been made.
The difference in cross section due to the choice of PDF's is 40 - 60\% at the highest mass.
MSTW2008NLO PDF's are favoured. However, it has to be noted that their determination included
the D\O\ inclusive jet production cross section measurement \cite{d0_incl_Jets_prl101_062001_2008}
which is based on the same data set as this dijet measurement. A further difference is that the 
MSTW2008NLO PDF's exclude Tevatron data taken before 2000 in contrast to the CTEQ6.6 PDF's.

\subsection{Inclusive 3-jet invariant mass cross section}
The inclusive 3-jet production cross section is measured~\cite{d0_xsec_3jetmass_Fermilab-PUB-11_173-E}
as a function of the invariant 3-jet mass based on an integrated luminosity of 0.7 fb$^{-1}$.
\begin{figure}[b]
\vspace*{-4ex}
\begin{center}
\includegraphics[width=13cm]{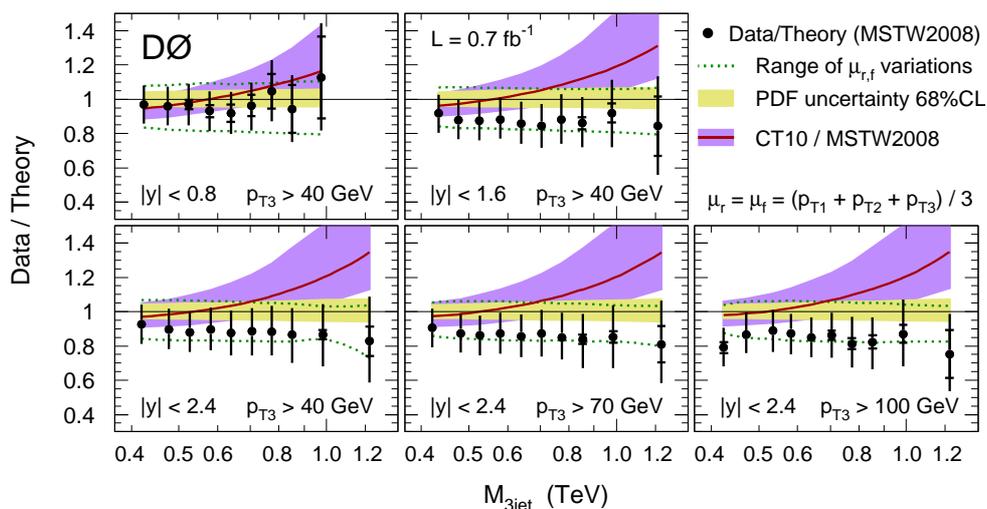}
\end{center}
\vspace*{-5.0ex}
\caption{ \label{3jetdiffxsec}
Data over theory ratios of the differential cross section 
$d\sigma_{\mbox{\scriptsize 3jet}}/dM_{\mbox{\scriptsize 3jet}}$ in the different rapidity regions and
for different third leading jet transverse momentum requirements. The pQCD predictions are given for two 
different PDF's.}
\end{figure}
Recent measurements of inclusive jet and dijet production in $p\bar{p}$ collisions at a centre 
of mass energy of $\sqrt{s}=1.96$~TeV 
\cite{cdf_prd75_092006_2007}\cite{cdf_prd78_052006_2008}\cite{cdf_prd79_119902_2009}\cite{d0_incl_Jets_prl101_062001_2008}\cite{d0_prl103_191803_2009}\cite{d0_plb693_531_2010}
have been used to determine $\alpha_s$ \cite{prd80_111107_2009} and the proton PDF's and to set 
limits on a number of physics models beyond the standard model \cite{Martin09}\cite{Lai10}\cite{Ball11}.
This shows the successful pQCD description of observables which are directly sensitive to matrix 
elements of ${\cal{O}}(\alpha_s^2)$. Testing pQCD at higher orders of $\alpha_s$ requires measuring 
cross sections of higher jet multiplicities. The three jet cross section is directly sensitive to 
matrix elements of ${\cal{O}}(\alpha_s^3)$ while the PDF sensitivity is similar.
Due to the fact that the pQCD calculations are available to next-to-leading order (NLO) in $\alpha_s$
\cite{Kilgore99}\cite{Kilgore97}\cite{nlojet}\cite{nlojet02}, the three jet cross section can be used 
for precision phenomenology, e.g. the simultaneous determination of $\alpha_s$ and PDF's from 
experimental data. Together with the inclusive jet and dijet cross sections the results for $\alpha_s$
and the PDF's get then partially decorrelated.

\begin{figure}[b]
\vspace*{-2ex}
\begin{center}
\includegraphics[width=13.0cm]{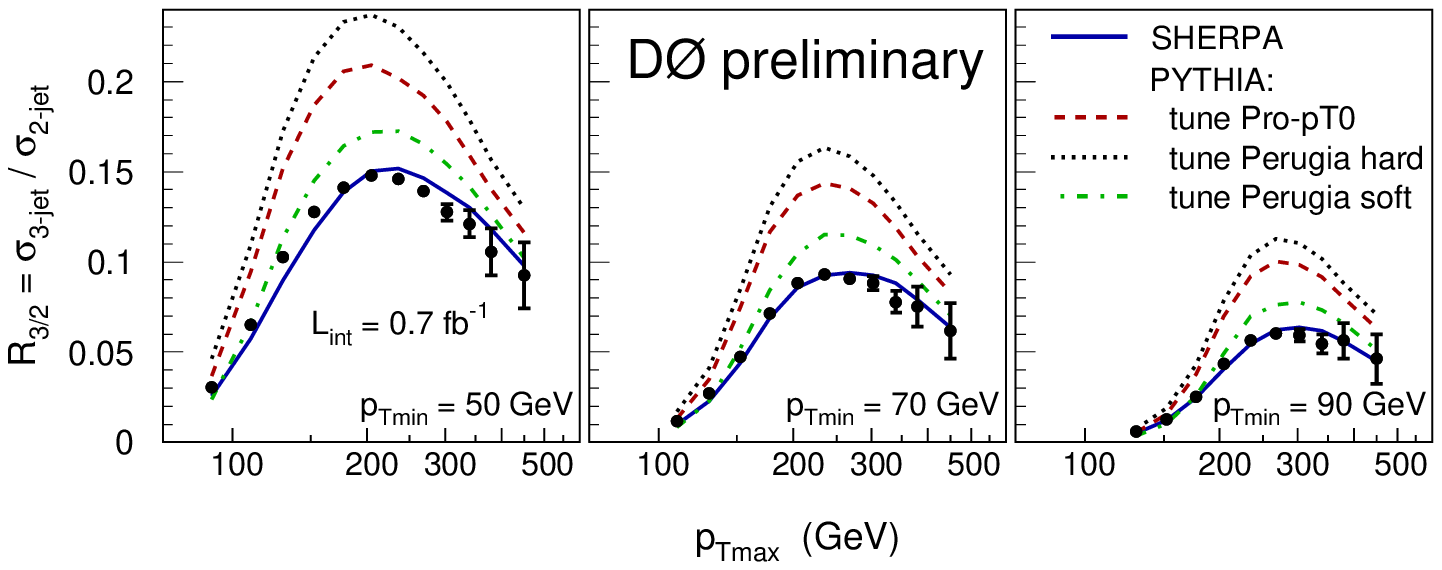}
\includegraphics[width=13.0cm]{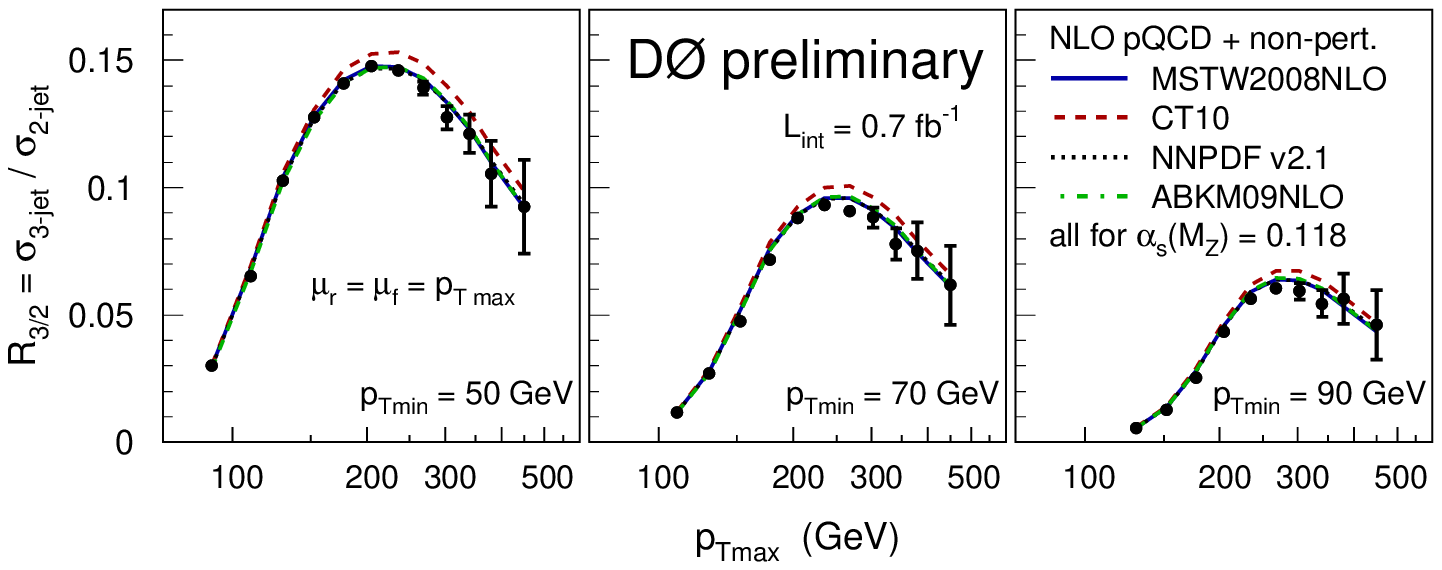}
\end{center}
\vspace*{-5ex}
\caption{ \label{R32}
Trijet inclusive over dijet inclusive cross section ratio as a function of the 
leading transverse momentum jet in comparison to various predictions of SHERPA and PYTHIA (top)
and next-to-leading order pQCD calculations (bottom).}
\end{figure}

The leading transverse momentum jet has to fulfil $p_T>150$~GeV and the third leading transverse 
momentum jet $p_T>40$~GeV. The leading jets 
are restricted to $|y|<0.8$, $|y|<1.6$, $|y|<2.4$ in three 
different measurements. Two additional measurements are made for a third leading jet $p_T>70$~GeV
and $p_T>100$~GeV in the rapidity range of $|y|<2.4$. 
The cone jet algorithm~\cite{run2cone} with radius $R=0.7$ is used and all pairs of the three 
leading jets are required to be spatially separated by $\Delta R>1.4 (=2\cdot R)$. 
Dominating systematic uncertainty on the cross section is given by the Jet Energy Scale (JES)
with 10 - 15\% for $|y|<0.8$ and 10 - 30\% for $|y|<2.4$. 
The $p_T$ resolution for jets is about 15\% at 40~GeV, decreasing to below 10\% at 400~GeV.
The corresponding uncertainty on the cross section amounts to 1 - 5\%.
The uncertainty on the luminosity contributes with 6.1\%.
In Fig.~\ref{3jetdiffxsec} the data over theory prediction ratios of the differential cross section 
$d\sigma_{\mbox{\scriptsize 3jet}}/dM_{\mbox{\scriptsize 3jet}}$ are shown in the different rapidity
regions and for different third jet transverse momentum thresholds. The MSTW2008NLO PDF uncertainties 
corresponding to the 68\% C.L. are shown by the light band. Normalised to MSTW2008NLO, 
the CT10 PDF's\cite{Lai10}
with corresponding value of $\alpha_s(M_Z)=0.118$ predict a different shape with increasing 
discrepancy at higher invariant masses.

\subsection{Ratio of inclusive trijet/dijet production $R_{3/2}$}
The ratio $R_{3/2}$ of inclusive trijet over dijet production is measured~\cite{d0_xsec_R32_confNote6032} 
as a function of the leading jet transverse momentum $p_T^{\max}$ based on an integrated 
luminosity of 0.7 fb$^{-1}$.
Jet production is sensitive to the dynamics of the fundamental interaction and the partonic structure
of the initial state hadrons. Measurements dedicated to investigate the interaction dynamics are 
preferentially based on observables which are insensitive to the PDF's. Such observables can be 
constructed as ratios of cross sections for which the PDF sensitivity cancels out. The ratio $R_{3/2}$
of inclusive trijet over inclusive dijet production is such an observable. In the $R_{3/2}$ measurement
reported here $\alpha_s$ is being probed up to jet transverse momenta of 500~GeV. 
Jets are defined by the cone jet algorithm \cite{run2cone} with radius $R=0.7$ and all pairs of 
$n$-leading jets ($n=2,3$) are required to be separated by $\Delta R>2\cdot R$. The cross section ratio
is determined in three bins of jet transverse momentum thresholds $p_T^{\min}>50$, $70$, $90$~GeV. 
The leading transverse momentum jet has to exceed the $n$-th leading jet by 
$p_T^{\max}>p_T^{\min}+30$~GeV to ensure sufficient phase space for the second and third leading jet
such that corrections due to the experimental $p_T$ resolution remain small.
In Fig. \ref{R32} (top) the cross section ratio is compared to the event generators SHERPA 1.1.3
\cite{sherpa} and PYTHIA 6.419 \cite{pythia6419} for three different tunes using a $p_T$-ordered 
parton shower. The MSTW2008LO PDF's have been deployed for the predictions of both event generators.
While SHERPA with leading order matrix elements for 2-, 3- and 4-jet production matched to parton shower
describes the data well, PYTHIA with tunes using the $p_T$-ordered parton shower is not able to 
describe the data. PYTHIA making use of the tune BW with $Q^2$-ordered parton shower is able to 
describe the data but a 
former D\O\ measurement of dijet azimuthal decorrelations \cite{d0_prl94_221801_2005} can not be 
described by the BW tune. As can be seen in Fig. \ref{R32} (bottom) the prediction of pQCD in 
next-to-leading order is describing the data very well. PYTHIA predicts non-perturbative corrections 
due to hadronisation and underlying event of 2.5\% $\pm$ 0.5\%. Therefore the discrepancies
of various tunes can not be explained by poor parameter 
choices for the hadronisation and/or 
underlying event models. It can only be explained by parameters which affect the perturbative physics
as implemented in the parton shower or by a fundamental limitation of the model itself.

\section{Vector boson plus jet production} \label{vec+jets}
Measurements of vector boson plus jet production provide fundamental tests of pQCD in particular at
high momentum scales. Furthermore these kind of processes can constitute the dominant background in 
measurements of single top quark and $t\bar{t}$ production as well as in searches for the Standard 
Model (SM) Higgs boson and for physics beyond the SM.
In particular $Z$ boson plus $b$ jet production is an important background for associated 
Standard Model Higgs boson production ($ZH\rightarrow Zb\bar{b}$) and also supersymmetric partners 
of $b$ quarks. Furthermore the $b$ quark density of the proton can be probed.

\subsection{Inclusive $Z$ boson cross section ratio $\sigma(p\bar{p}\rightarrow Z + b + X)/\sigma(p\bar{p}\rightarrow Z + j + X)$}
The ratio of the inclusive production of a $Z$ boson accompanied by a $b$ jet over the inclusive 
production of a $Z$ boson accompanied by a light flavoured jet 
$\sigma(p\bar{p}\rightarrow Z + b + X)/\sigma(p\bar{p}\rightarrow Z +j + X)$ is 
measured~\cite{d0_incl_xsec_ZboverZj_ratio} as a function of the leading jet transverse momentum 
based on an integrated luminosity of 4.2 fb$^{-1}$.
The measurement of the cross section ratio benefits from cancellations of many systematic 
uncertainties.

Selected events are required to pass at least one of the single electron or single muon triggers.
A $Z$ boson candidate with an 
invariant dilepton mass of $70 < m_{\ell\ell} < 110$~GeV has to be reconstructed. Electrons
(muons) have to have a transverse momentum of $p_T>15$~GeV ($p_T>10$~GeV). Electron candidates are 
built from isolated energy deposits in the electromagnetic calorimeter with a shower shape consistent
to that expected from electrons. Combined tracking and calorimeter isolation requirements are 
applied to the muon candidates.
At least one jet with $p_T>20$~GeV and further jets with $p_T>15$~GeV, making use of the
cone jet algorithm \cite{run2cone} with cone radius $R=0.5$ are required. Jets with $b$ content have a 
different energy response and receive an additional average energy correction of about 6\%. 
The $b$-tagging algorithm is based on a neural network exploiting the longer lifetime of
$b$-flavoured hadrons which results in tracks with sizeable impact parameters and secondary vertices. 
The necessary tracks associated to jets have to exceed a transverse momentum of 0.5~GeV and 1.0~GeV 
for the leading track. At least one of the jets has to be $b$-tagged.
The tagging efficiency for $b$ jets and light jets are parameterised as functions of jet 
$p_T$ and pseudorapidity $\eta$. They amount to about 58\% and 2\% respectively.
To further separate $b$ jets from $c$ and light jets a discriminant based on the secondary vertex 
mass $M_{SV}$ and the jet lifetime impact parameter (JLIP) is introduced.
Events with missing transverse energy $E\!\!\!\!/\;_T>60$~GeV are rejected to 
suppress background from $t\bar{t}$ production. 
The dominant background to $Z$ + jet production arises from multijet events with jets misreconstructed
as leptons, especially in the $ee$ channel. Smaller background contributions arise from $t\bar{t}$
and diboson ($ZW$, $ZZ$, $WW$) production.
A binned maximum likelihood fit to the discriminant distribution in data using a linear combination
of light, $c$ and $b$ flavoured jet templates from simulation is performed to determine the flavour 
fractions. Several experimental uncertainties cancel out in the measurement of the cross section ratio
$\sigma(Z + b + X)/\sigma(Z + j + X)$, including the uncertainties on the luminosity, trigger,
lepton identification and some jet identification efficiencies. The two largest remaining sources 
of systematic uncertainty are due to the discriminant efficiency and the shape of the discriminant 
templates. Other important sources of uncertainty are the $b$-tagging efficiency with 2.4\%, the $b$ 
jet energy scale with 2\% and reconstruction efficiency with 3.2\%.
The measurement of the ratio yields
$\frac{\sigma(Z+ b\; \mbox{\scriptsize jet})}{\sigma(Z + \mbox{\scriptsize jet})} = 0.0193 \pm 0.0022(\mbox{stat}) \pm 0.0015(\mbox{syst})$ consistent with the ratios obtained separately in the two lepton channels and the NLO prediction
of MCFM~\cite{mcfm} of $0.0192\pm 0.0022$ for renormalisation and factorisation scales
$\mu^2_R=\mu^2_F=m^2_Z$ making use of the MSTW2008~\cite{mstw2008} PDF's.

\subsection{Inclusive $W (\rightarrow e\nu_e)$ plus jets production rates}
\begin{figure}[t]
\vspace*{-3ex}
\includegraphics[width=8.2cm]{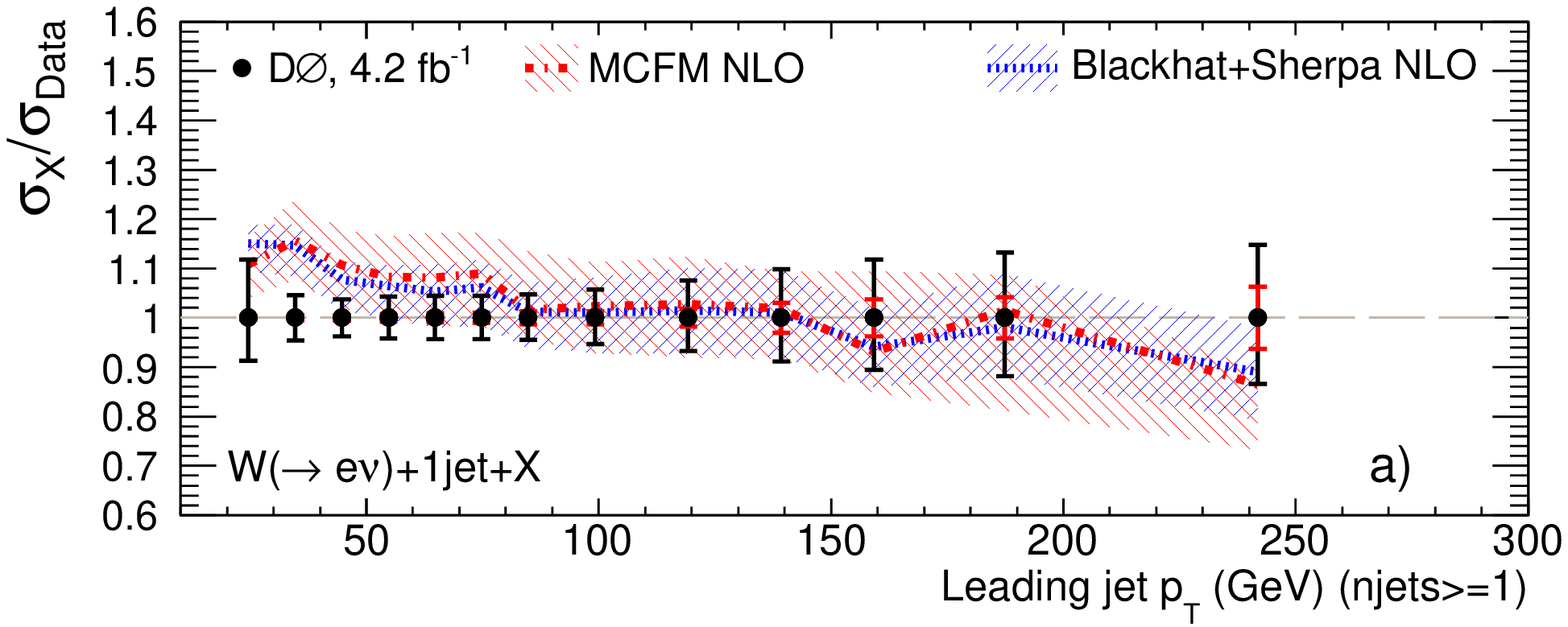}
\includegraphics[width=8.2cm]{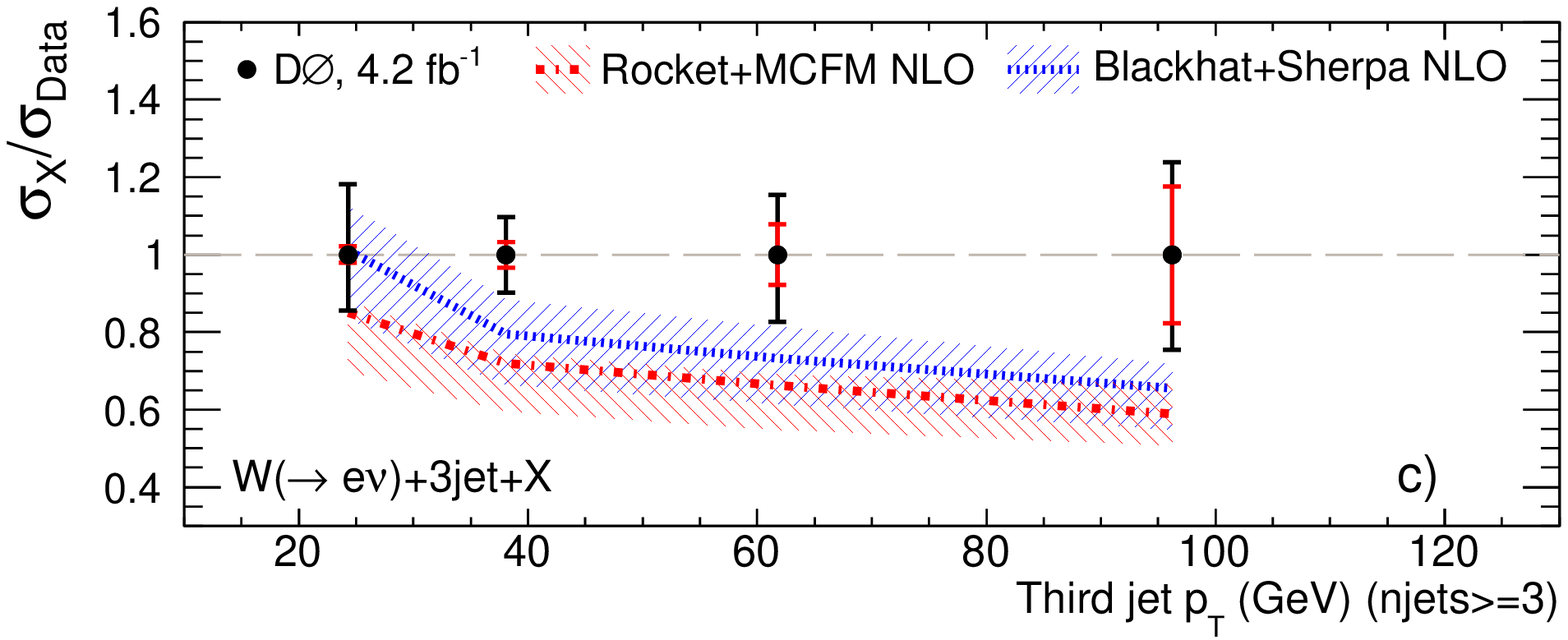}
\includegraphics[width=8.2cm]{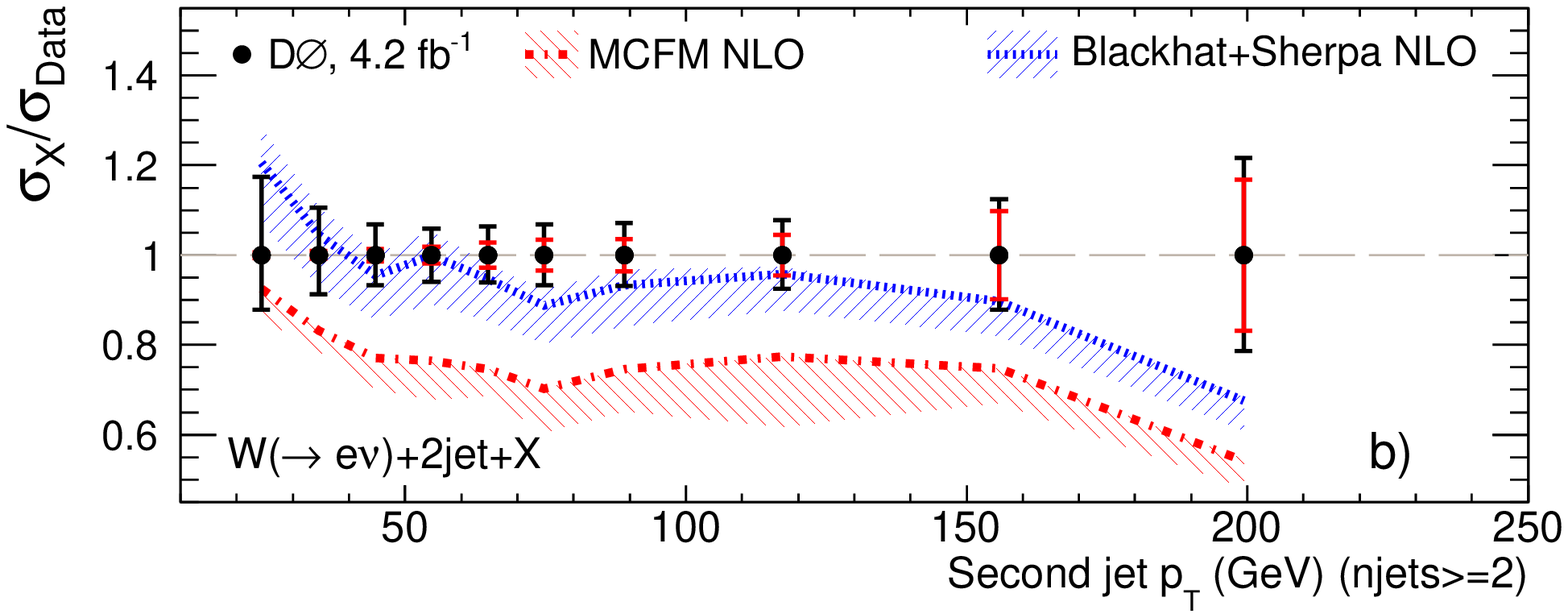}
\includegraphics[width=8.2cm]{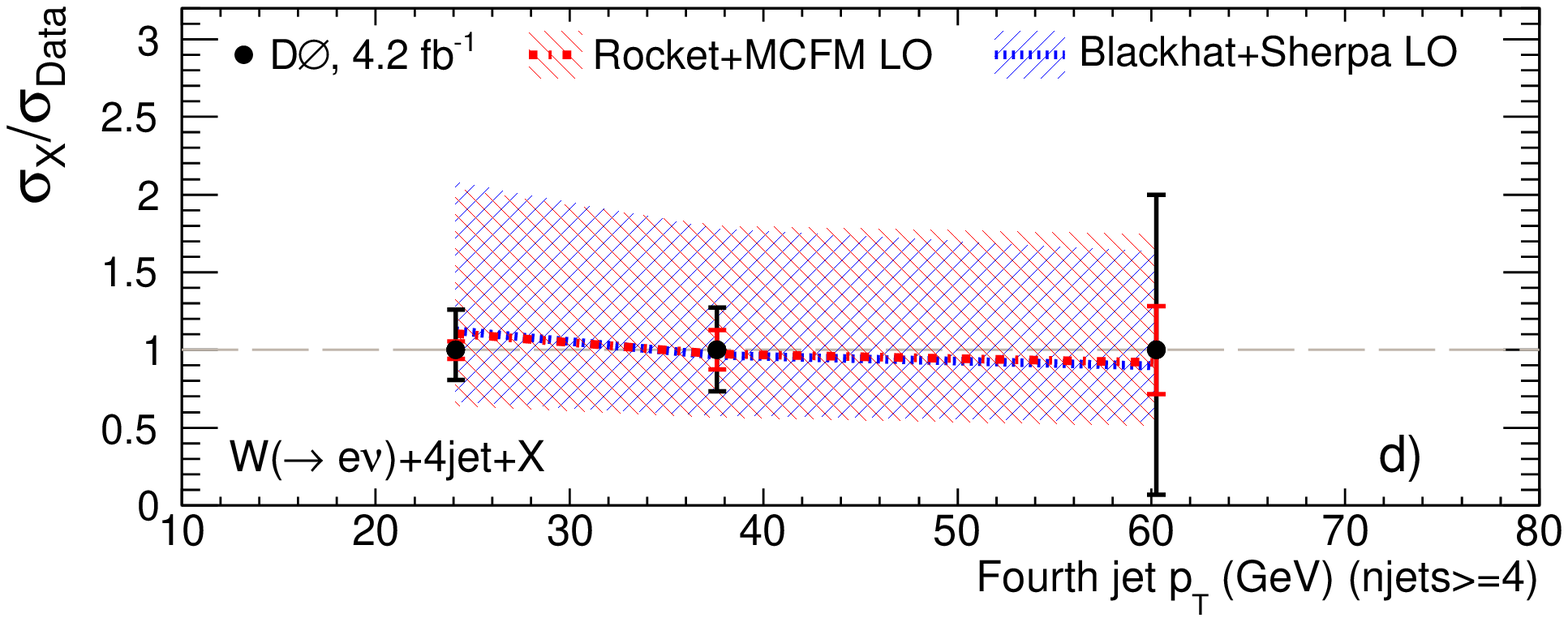}
\vspace*{-5ex}
\caption{ \label{Wjets}
Ratio of pQCD predictions to measured differential cross sections for the $p_T$ of the $n$-th leading
jet in $W + n$ jet production (left: $n=1$ (top), $n=2$ (bottom). Right: $n=3$ (top), $n=4$ (bottom)).
The shaded areas indicate the theoretical uncertainties due to variations of the factorisation and 
renormalisation scales.
}
\end{figure}
The inclusive $W$ boson plus jets production rates are measured~\cite{d0Note_WenuJets} in dependence
of the inclusive number of jets based on an integrated luminosity of 4.2 fb$^{-1}$.
The results are corrected to the hadronic final state in deconvolving the effects of finite 
detector resolution, detector response, acceptance and efficiencies.
Jets are identified with the D\O\ midpoint cone algorithm~\cite{run2cone} with cone radius $R=0.5$.
Jet energy corrections are determined by means of the transverse momentum imbalance in
$\gamma$ + jet events, where the electromagnetic calorimeter response is calibrated using
$Z/\gamma^*\rightarrow e^+e^-$ events. The missing transverse energy $E\!\!\!\!/\;_T$ is corrected 
for the presence of any muon in the event. Central electrons with $p_T>15$~GeV, missing transverse
energy $E\!\!\!\!/\;_T>20$~GeV,
transverse mass of the $W$ boson candidate $M_T^W>40$~GeV
and jet transverse  momenta of $p_T>20$~GeV are required. Furthermore, the distance between electrons
and the nearest jet has to satisfy $\Delta R>0.5$ in azimuthal angle $\phi$ and pseudorapidity $\eta$ 
space. After this selection there are backgrounds from $Z$ + jets, $t\bar{t}$, diboson, single top 
quark and multijet production. $W/Z$+jets and $t\bar{t}$ processes are simulated with ALPGEN v2.11~\cite{alpgen211}, interfaced to PYTHIA v6.403~\cite{pythia6403} for the simulation of initial and final 
state radiation and for the hadronisation. The PYTHIA event generator is used to simulate diboson 
production. Single top quark production is simulated via the COMPHEP~\cite{comphep} generator
interfaced to PYTHIA. The cross sections for $W/Z$ + jet production used for acceptance and smearing 
corrections are simulated by ALPGEN, corrected with a constant factor to match the inclusive
$W/Z$ + jet cross section calculated at NLO~\cite{campbell02}\cite{rainwater03}. 
Additional corrections are applied to $W/Z$ + heavy flavour jet production to match the prediction
of NLO calculations. Multijet background is determined by means of a data-driven 
method~\cite{d0_prd76_092007_2007} in which data in a control region orthogonal to the data 
selection is used to determine shape and overall normalisation of the multijet contribution.
After background subtraction the data are unfolded, including an unfolding bias correction of the 
order 0.5 - 2\%. All differential cross section measurements are normalised to the measured 
inclusive $W$ boson cross section. The total inclusive $W$ boson cross section is measured to be
$\sigma_W = 1132\pm 1(\mbox{stat})^{+43}_{-84}(\mbox{syst})+69(\mbox{lumi})$~pb with 
$W\rightarrow\tau\nu_{\tau}\rightarrow e\nu_{e}\nu_{\tau}$ contributing as signal. A recent publication
\cite{berger11} of $W+4$ jet production at NLO for $pp$ collisions at $\sqrt{s}=7$~TeV is available.
These predictions are not available for the Tevatron. Comparisons with theory are therefore limited to
LO for $W+4$ jet production. For the predictions up to three partons in the final state the
partonic final state generators BLACKHAT~\cite{blackhat}, interfaced to SHERPA~\cite{sherpa} and 
ROCKET~\cite{rocket} interfaced to MCFM~\cite{mcfm} are used.
BLACKHAT + SHERPA deploys the renormalisation and factorisation scale 
$\mu=\mu_R=\mu_F=\frac{1}{2}\hat{H}_T$ where $\hat{H}_T$ is the scalar sum of parton and lepton
transverse energies. The choice made by the ROCKET + MCFM authors is 
$\mu=\sqrt{M_W^2+\frac{1}{4}(\sum p^{\mbox{\scriptsize jet}})^2}$ in the $n=2,3$ and 4 jet bins.
In the 1 jet bin the factor $1/4$ is omitted. 
The PDF's used for the theory calculations are MSTW2008~\cite{mstw2008}. 
Hadronisation corrections are obtained with SHERPA and the CTEQ6.6 PDF's.
Fig. \ref{Wjets} shows ratios of the pQCD predictions to the differential cross sections for 
$W$ boson plus jet production in the jet multiplicity bins $n=1-4$ as a function of the $n$-th
jet transverse momentum respectively. The BLACKHAT + SHERPA predictions are in very good agreement
with the data everywhere except at small jet $p_T$. The ROCKET + MCFM predictions are significantly 
below the data. In particular the difference in the $W$ + 2 jet bin indicates that the scale 
uncertainties are larger than determined. The data offers here guidance for good choices of the 
renormalisation and factorisation scales.

\section{Direct photon production} \label{directPhotons}
Direct photons come unaltered from the hard subprocess and provide therefore a direct probe of the 
hard scattering dynamics. The energy calibration of photons is better than the energy calibration of jets.
Direct photons are also important background to many physics processes. At the same time the 
cross section of multijet production dominates. Therefore the understanding of the QCD production 
mechanism is indispensable in search for new physics. 
 
Photon candidates are given by electromagnetic showers in the calorimeter. By means of shower profile
and geometric isolation criteria they are distinguished from background, such as neutral mesons and
jets with high electromagnetic fraction. No associated track has to be found and jets have to be 
separated by a geometrical distance of $\Delta R>0.7$ in ($\eta,\phi$)-space. Multiple preshower hits 
help to distinguish prompt photons from $\pi^0\rightarrow 2\gamma$ decays.

\begin{figure}[b]
\vspace*{-5.0ex}
\begin{center}
\includegraphics[width=14cm]{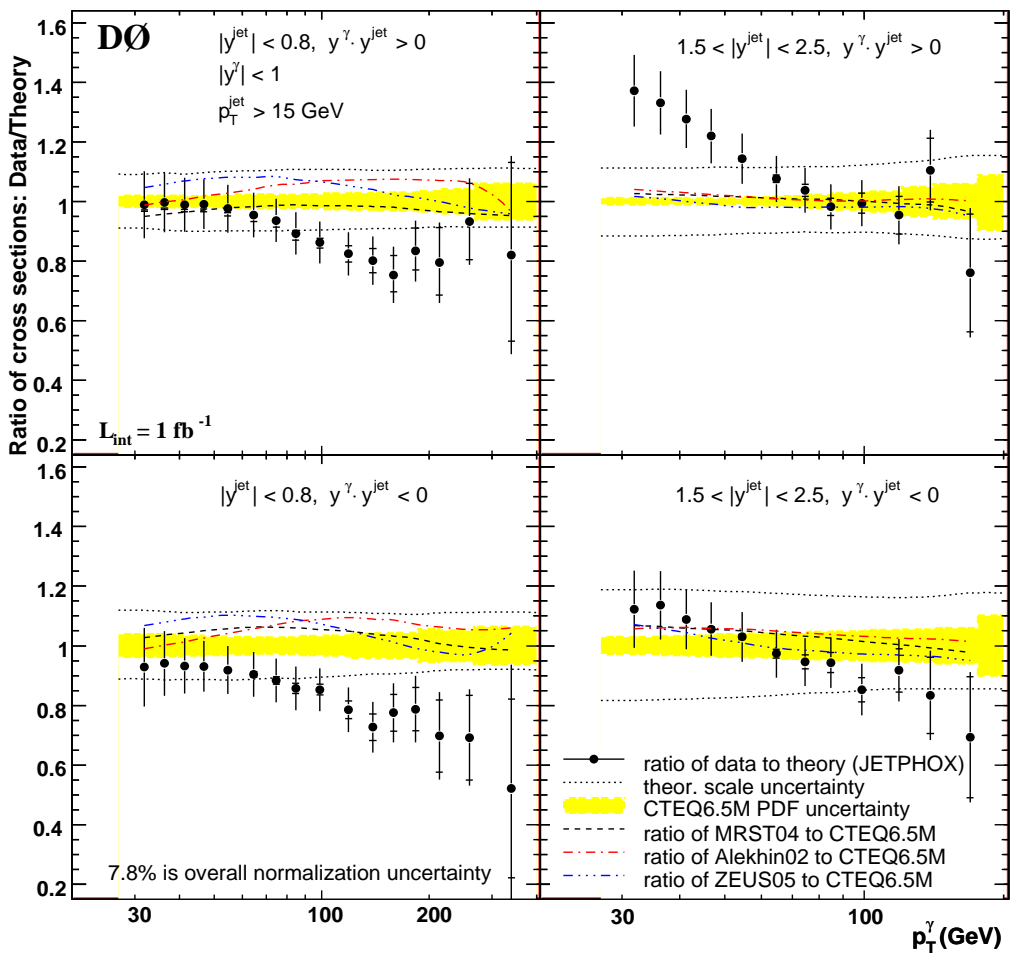}
\end{center}
\vspace*{-4ex}
\caption{ \label{gammajet}
Ratio of measured triple differential cross section for same/opposite photon and jet rapidities
and  central or forward leading jets as a function of the photon transverse momentum.
}
\end{figure}

At low photon transverse momentum ($p_T^{\gamma} < 120$~GeV) 
photon + jet production through 
Compton scattering $qg\rightarrow q\gamma$ dominates. The requirement of isolated photons reduces the 
contribution of photon fragmentation of quarks.
Photon + jet production can be exploited to probe PDF's at low $x$ (down to $x\simeq 0.007$ 
at the Tevatron), where quarks are constrained by HERA data and therefore one is sensitive 
to the gluon density. Finally pQCD can be probed at NLO, including soft gluon resummation 
and models of gluon radiation.

\subsection{Inclusive photon plus jet production}

The inclusive photon plus jet production is measured~\cite{d0_gamma+jets_plb666_435_2008} 
as a function of the photon transverse momentum based on an integrated luminosity of 1.0 fb$^{-1}$.
An isolated central photon with $p_{T} >30$~GeV is required to pass a neural network background 
discrimination based on tracker and calorimeter information. Jets above a transverse momentum of
$p_{T} >15$~GeV in the two rapidity intervals $|y|<0.8$ (central) and $1.5<|y|<2.5$ (forward) are 
selected. Four regions are distinguished, depending on if the photon and jet have the same or opposite
rapidity sign
($y^{\gamma}\cdot y^{\mbox{\scriptsize jet}} <0$ or $>0$)
and depending on the central or forward location of the leading jet. 
The triple differential cross section 
as a function of the photon transverse momentum and rapidity and the jet rapidity 
$d^3\sigma/dp_T^{\gamma}dy^{\gamma}dy^{\mbox{\scriptsize jet}}$ is measured. The NLO prediction
is given by JETPHOX~\cite{jetphox_a}\cite{jetphox_b} making use of CTEQ6.5~\cite{cteq6.5} PDF's 
and BFG photon fragmentation functions~\cite{bfg}. In Fig. \ref{gammajet} the triple 
differential cross section is shown for the four different phase space regions, depending on the
same/opposite photon jet hemisphere and the central/forward leading jet location as a function 
of the photon transverse momentum. The theory is not able to describe the data in the whole
measured range. The structure is similar to previous observations of UA2~\cite{ua2_photon}, 
CDF~\cite{cdf_photon} and D\O~\cite{d0_photon}\cite{d0_photon_b}.
The cross sections as a function of the photon $p_T$ can also be arranged as ratios over two different
regions (out of the four possible regions) giving rise to six different ratios which benefit from 
reduced systematics. The shapes of the measured cross section ratios in data are qualitatively 
reproduced by the theory in general, but quantitative disagreement is observed for some kinematic 
regions. This disagreement holds in particular for central jets over same sign rapidity forward jets.

\subsection{Inclusive photon plus heavy flavour jet production}
The inclusive photon plus heavy flavour jet production is measured~\cite{d0_gamma_bjet_prl102_192002_2009}
as a function of the photon transverse momentum based on an integrated luminosity of 1.0 fb$^{-1}$.
The QCD Compton photo production process dominates for photon transverse momenta of 
$p_T\sim 90$ $(120)$~GeV for $b$ ($c$) quark flavour. Since the outgoing quark flavour is for this 
production mechanism equal to the incoming quark flavour, constraints can be set on the heavy flavour 
content of the PDF's. At larger photon $p_T$ the quark anti-quark annihilation process with a photon
and a gluon dominates, where the gluon splitting into a charge conjugated heavy flavour quark pair 
enters into the considered data sample. An isolated central photon with transverse momentum of 
$p_T>30$~GeV is required together with central jets (cone radius $R=0.5$) above a transverse momentum 
of 15~GeV. A neural network is applied to the jets to enhance the heavy flavour content of the selected
data sample. Templates for different flavour contents are built by means of a function of jet track 
probabilities. The distributions of these functions are fitted to the shape of the data to extract the 
flavour composition. The measurement is done separately in the two kinematic regions given by
same and opposite photon and jet rapidities. The cross sections for the $b$ and $c$ quark flavours in the 
two kinematic regions are measured as a function of the photon transverse momentum.
Comparison to theory~\cite{owens} reveals disagreement for $\gamma$ + $c$ jet production for photon 
transverse momenta above 70~GeV.

\begin{figure}[b]
\includegraphics[width=5.4cm]{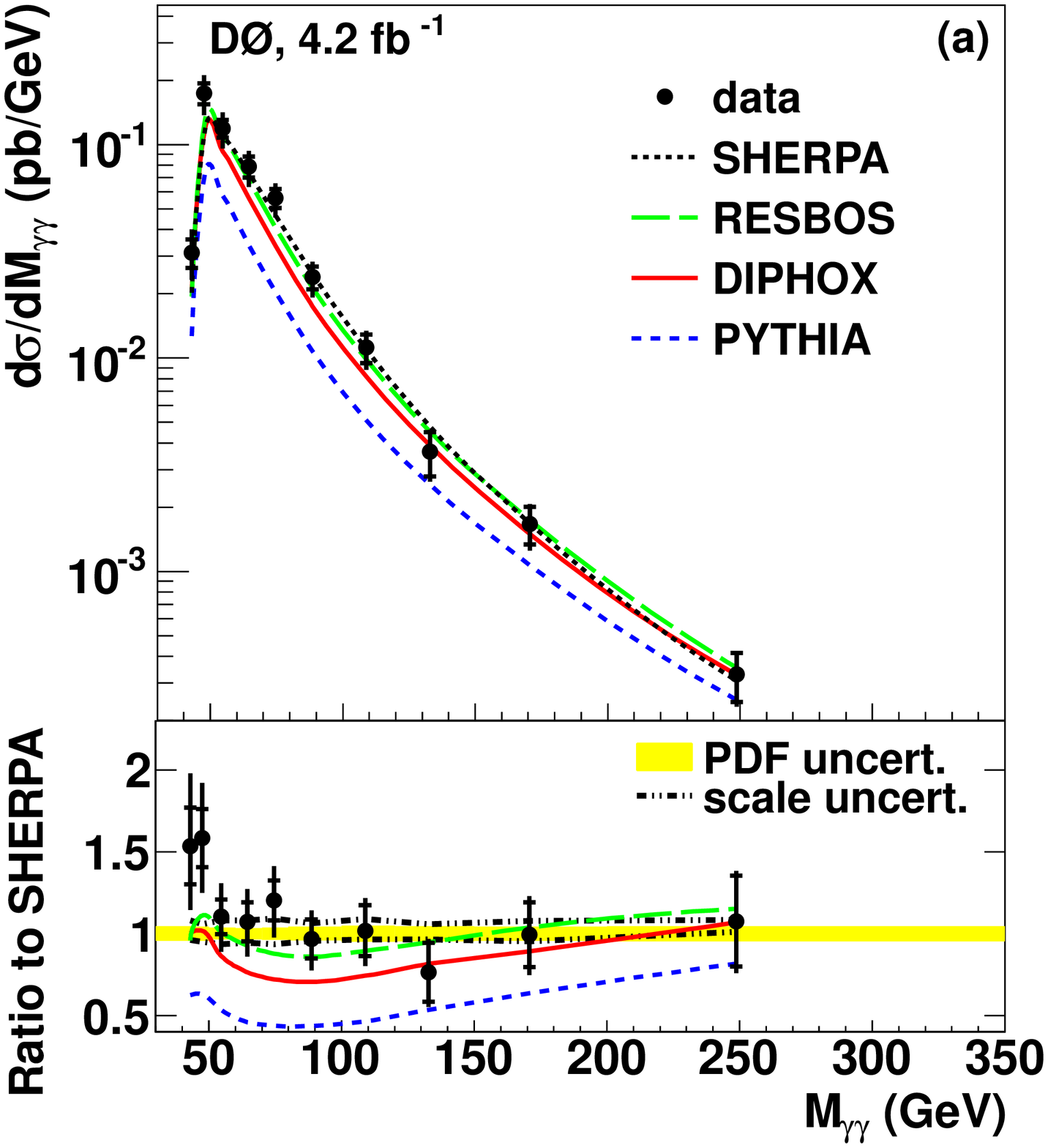}
\includegraphics[width=5.4cm]{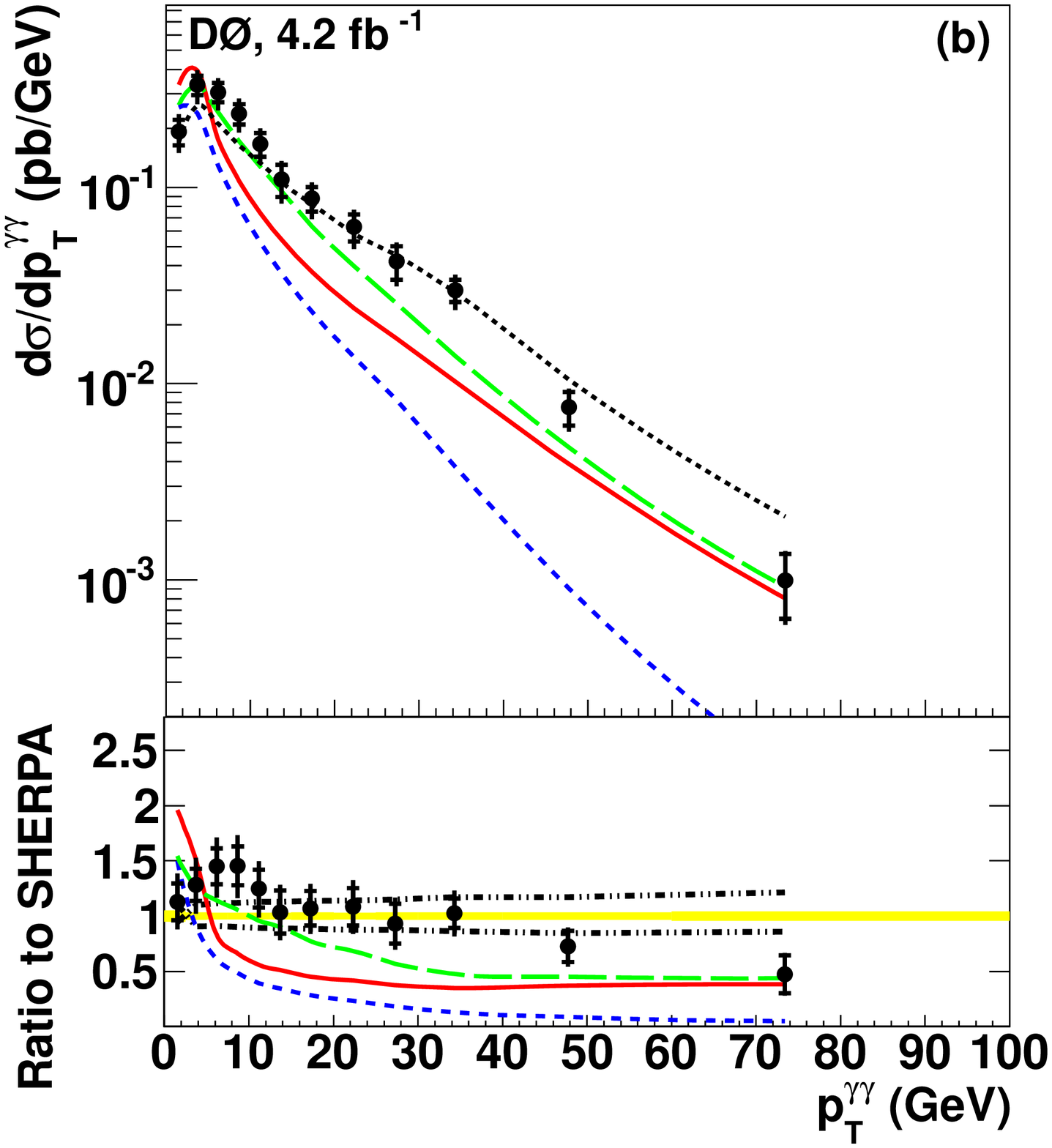}
\includegraphics[width=5.4cm]{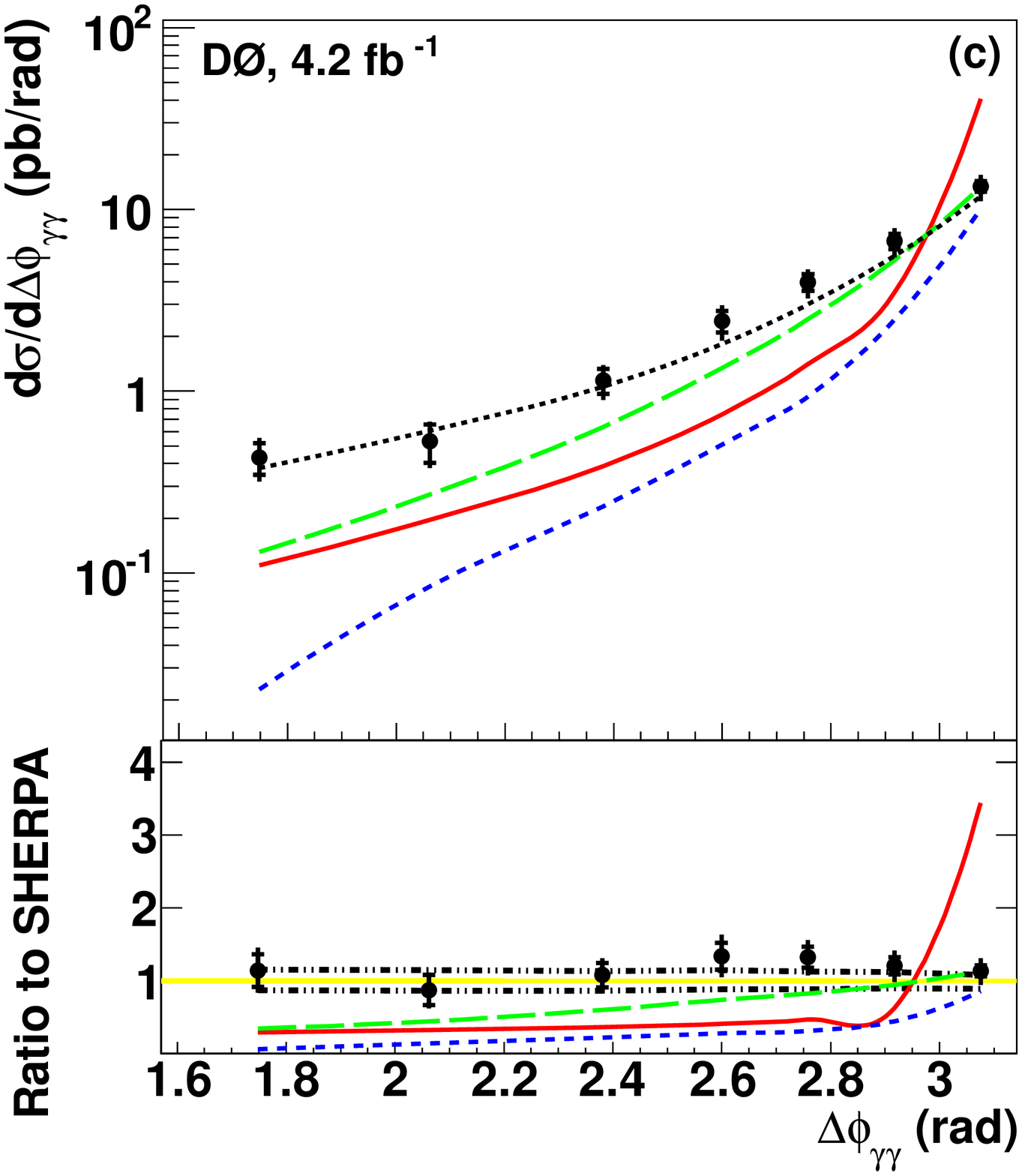}
\vspace*{-7ex}
\caption{ \label{diphoton}
Differential diphoton production cross section as a function of the invariant diphoton mass
$m_{\gamma\gamma}$ (a), the transverse diphoton momentum $p_T^{\gamma\gamma}$ (b) and the 
azimuthal angle between the photons $\Delta\phi_{\gamma\gamma}$ (c) in comparison to theory and 
various models.
}
\end{figure}

\subsection{Direct photon pair production}
The direct photon pair production cross section is measured~\cite{d0_2gamma_plb690_108_2010} 
as function of the invariant diphoton mass $M_{\gamma\gamma}$, 
the diphoton transverse momentum $p_T^{\gamma\gamma}$,
the azimuthal angle between the two 
photons $\Delta\phi_{\gamma\gamma}$ and the cosine of the diphoton scattering angle in the diphoton 
rest frame $\cos\theta^*=\tanh [(\eta_1-\eta_2)/2]$, based on an integrated luminosity of 4.2 fb$^{-1}$.
Direct photon pair production with large invariant mass $m_{\gamma\gamma}$ constitutes a 
large and irreducible background to searches for the Higgs boson decaying into a pair of photons
at hadron colliders. It is also a significant background in searches for new phenomena, such as new 
heavy resonances~\cite{mrenna01}, extra spatial dimensions~\cite{kumar09}, or cascade decays of heavy 
new particles~\cite{guidice99}.
Double photon production is also interesting to check the validity of pQCD and soft gluon resummation
implemented in theoretical calculations. Background contributions from single and double photon 
fragmentation are suppressed by photon isolation and invariant $m_{\gamma\gamma}$ requirements, which 
entails smaller theoretical uncertainties. 
Events with two central isolated photons of transverse momentum $p_T^{\gamma_1}>21$~GeV and 
$p_T^{\gamma_2}>20$~GeV are selected with unbalanced thresholds to ensure non-vanishing phase space 
at NLO. The invariant mass is required to exceed the transverse momentum of the diphoton system 
$m_{\gamma\gamma}>p_T^{\gamma\gamma}$. The two photons have to be separated by the distance 
$\Delta R>0.4$ in  ($\eta,\phi$)-space.
The cross sections are measured differentially as a function of $m_{\gamma\gamma}$, the diphoton
transverse momentum $p_T^{\gamma\gamma}$, the azimuthal angle between the photons 
$\Delta\phi_{\gamma\gamma}$ (see Fig.~\ref{diphoton} for a data comparison to theory and various models) 
and the cosine of the scattering angle 
$\cos\theta^*$, where holds $p_T^{\gamma_1}>p_T^{\gamma_2}$.
The shapes of the $p_T^{\gamma\gamma}$ and $\Delta\phi_{\gamma\gamma}$ distributions are mostly 
affected by initial state gluon radiation and fragmentation effects.
Furthermore, the $m_{\gamma\gamma}$ spectrum is in particular sensitive to potential contributions 
from new phenomena. The $\cos\theta^*$ distribution probes PDF effects and the angular momentum of 
the final state. The measured cross sections are compared to theoretical predictions from
SHERPA~\cite{sherpa}, RESBOS~\cite{resbos}\cite{resbos_b}, DIPHOX~\cite{diphox} and PYTHIA~\cite{pythia6420}.
Both, RESBOS and DIPHOX provide pQCD NLO predictions, however the $gg\rightarrow\gamma\gamma$
contribution is implemented only at leading order (LO) in DIPHOX. The explicit 
parton-to-photon fragmentation functions are included in DIPHOX at NLO, while in RESBOS a function 
approximating rates from NLO fragmentation diagrams is implemented. Furthermore only in RESBOS the 
effect of soft and collinear initial state gluon emission is resummed to all orders.
The normalisation uncertainty on the measured cross section amounts to 7.4\% due to uncertainties
on the luminosity of 6.1\% and on the selection criteria of 4.3\%. 
PDF uncertainties are determined by means of DIPHOX, making use of the CTEQ6.6M~\cite{cteq6.6} PDF's.
The renormalisation, factorisation and fragmentation scales are set to 
$\mu_R=\mu_F=\mu_f=m_{\gamma\gamma}$. Studies performed using DIPHOX indicate that the contribution
to the overall cross section from one- and two-photon fragmentation processes does not exceed 16\%
and significantly drops at large $m_{\gamma\gamma}$, $p_T^{\gamma\gamma}$ and small 
$\Delta\phi_{\gamma\gamma}$ to 1 - 3\%. None of the theoretical predictions considered is able to 
describe the data well in all kinematic regions of the four variables. RESBOS shows the best 
agreement with data, although systematic discrepancies are observed at low $m_{\gamma\gamma}$,
high $p_T^{\gamma\gamma}$ and low $\Delta\phi_{\gamma\gamma}$. The large discrepancy between
RESBOS and DIPHOX in some regions of the phase space is due to the absence of all-order soft gluon
resummation and the LO approximation of the $gg\rightarrow\gamma\gamma$ contribution in DIPHOX.
Further insight can be obtained in considering double differential cross sections. Therefore the
differential cross sections as functions of $p_T^{\gamma\gamma}$, $\Delta\phi_{\gamma\gamma}$ 
and $|\cos\theta^*|$ are measured in three invariant
$m_{\gamma\gamma}$ mass bins 30 - 50~GeV, 50 - 80~GeV and 80 - 350~GeV.
The largest discrepancies between data and RESBOS are found for each kinematic variable in the lowest 
$m_{\gamma\gamma}<50$~GeV region. This is the region, where the contribution of 
$gg\rightarrow\gamma\gamma$ is expected to be largest.
Finally, the addition of NNLO corrections to RESBOS remains to be compared, once available.

\section{Minimum bias and double parton scattering} \label{MB+DPS}
The non-perturbative nature of low energy QCD processes makes it necessary to develop heuristic 
descriptions. Models that describe these soft processes are implemented in Monte Carlo event generators.
They contain many parameters that need to be tuned using experimental input. Most prominent input
observables are charged particle multiplicities and track mean transverse momenta.


Double Parton Scattering (DPS) may constitute a background for many rare processes of interest,
especially those with multijet final states.
The measurement of DPS provides complementary information about the proton 
structure in terms of the spatial distributions of partons. Furthermore it allows one to probe 
for parton-parton correlations and therefore to check for a possible impact on the PDF's.
Multiple Parton Interaction (MPI) models can be constrained and improved taking DPS measurements 
into account.

\subsection{$\phi$ and $\eta$ correlations in minimum bias events}
The azimuthal angle and pseudorapidity correlations in minimum bias events are 
measured~\cite{d0_etaPhiCorr_MB_confNote6054} based on a data set taken in the years 2002 to 2006.
Complementary to existing observables the azimuthal angle difference $\Delta\phi$ between the
highest transverse momentum track and any other track in a given event is considered here.
This observable has the advantage of being as independent as possible of the track fake rate and the
tracking efficiency. The general shape of the $\Delta\phi$ distribution consists of peaks at 0 and 
$\pi$ on top of a pedestal of certain height.

Events with more than one reconstructed proton anti-proton interaction per bunch crossing are 
selected and the single collision that fired the trigger is being identified. The remaining 
collisions of the same bunch crossing are regarded as minimum bias events. The triggers used are 
dimuon triggers. The primary vertices of these collisions are required to have at least five tracks
with transverse momentum above 0.5~GeV inside the geometrical tracker acceptance ($|\eta|<2$).
Long-lived particles, such as pions and $K_s$ are discriminated by means of a vertex $\chi^2$
requirement based on three dimensional track impact parameter significances.
Uncorrelated fake tracks and wrongly associated real tracks are distributed uniformly in $\Delta\phi$.
This flat background is eliminated in fitting the distribution by a polynomial and subtracting the 
height at minimum from the whole distribution.
The absolute level of the shape of the $\Delta\phi$ distribution is affected by the tracking efficiency
but the dependence is minimised by normalising the distribution to unit area and only considering the 
shape. As it turned out it has not been necessary to unfold the experimental result.
A further observable has been defined  by assigning the tracks into two pseudorapidity regions based on 
the sign of the leading tracks $\eta$. The remaining tracks belong then to the same or opposite
region depending on the sign of their $\eta$ respectively. The number of same sign tracks subtracted 
by the number of opposite sign tracks in a given $\Delta\phi$ bin, normalised to the sum of all such
expressions over all $\Delta\phi$ bins is then the final observable which is being considered as a 
function of $\Delta\phi$. The observables are compared to three PYTHIA tune and model implementations,
which are Rick Field's Tune A~\cite{TuneA01}, the Perugia 0 tune (P0)~\cite{P0tune09} and the
Generalised Area Law model of colour reconnections (GAL)~\cite{gal99}.
Tune A is historically significant and makes use of $Q^2$-ordered parton showers while the more recent 
tune P0 makes use of $p_T$-ordered showers. Both use the colour annealing model for colour reconnections.
GAL is based on the tune S0~\cite{S0_2005} but uses the Generalised Area Law colour reconnection model.
The different tunes show large differences for the considered observables which provide therefore
important additional information to be taken into account in tunes.

\begin{table}[t]
\vspace*{-2ex}
\begin{center}
\begin{tabular}{|c|c|c|c|c|} 
\hline
Model & $\rho(r)$ & $\sigma_{\mbox{\scriptsize eff}}$ & $R_{\mbox{\scriptsize rms}}$ &  $R_{\mbox{\scriptsize rms}}$ (fm) \\ \hline
Solid sphere & const., $r<r_p$ & $4\pi r_p^2/2.2$ & $\sqrt{3/5} r_p$ & $0.41\pm0.05$ \\
Gaussian & $e^{-r^2/2a^2}$ & $8\pi a^2$ & $\sqrt{a}$ & $0.44\pm0.05$ \\
Exponential & $e^{-r/b}$ & $28\pi b^2$ & $\sqrt{12}b$ & $0.47\pm0.06$ \\ \hline
\end{tabular}
\end{center}
\vspace*{-2ex}
\caption{ \label{spatialProton}
Parameters of different spatial parton density models calculated from $\sigma_{\mbox{\scriptsize eff}}$.
}
\end{table}

\subsection{Double parton scattering in $\gamma$ + 3 jet events}
The double parton scattering fraction is measured~\cite{d0_DPS_prd81_052012_2010} in dependence 
of the second leading jet transverse momentum based on an integrated luminosity of  1.0 fb$^{-1}$.
The cross section for DPS is given by 
$\sigma_{DP}= m\cdot\frac{\sigma_A\sigma_B}{2\sigma_{\mbox{\scriptsize eff}}}$, where $\sigma_A$
and $\sigma_B$ are the cross sections of the processes $A$ and $B$, $m$ is a constant permutation 
factor which corresponds to one in case of indistinguishable processes $A$ and $B$. It 
assumes the value $m=2$ in the case of distinguishable processes $A$ and $B$ as discussed 
here. $\sigma_{\mbox{\scriptsize eff}}$ is a process independent scaling parameter which gives a 
measure for the size of the effective interaction region.
The expression $\sigma_B/2\sigma_{\mbox{\scriptsize eff}}$ can be interpreted as the probability
of a second interaction $B$ given that a first one $A$ has taken place.

DPS is measured in $\gamma$ + 3 jet events which benefits from the better energy measurement of
the photon and a larger fraction of Double Parton (DP) events compared to 4-jet events~\cite{drees96}.
Events with an isolated high transverse momentum photon of $60<p_T<80$~GeV are selected to increase 
the photon purity. Furthermore this requirement ensures a clean separation of the jet produced in 
the same parton scattering and jets produced by a different scattering. The jet transverse momenta 
are corrected to the hadronic final state. The first jet in transverse momentum is required to
fulfil $p_T>25$~GeV and the second and third jet $p_T>15$~GeV. Jets are reconstructed with the 
iterative midpoint cone jet algorithm~\cite{run2cone} with cone size $R=0.7$. The main background is Single
Parton (SP) scattering. Also Double Interaction (DI) events with two scatterings of two different 
$p\bar{p}$ collisions in the same bunch crossing (referred to as pile-up) have to be considered.
They can be distinguished by the separate collision vertex.
The contribution from single and double diffraction events represents 
$< 1\%$
of the total dijet cross section. 
The $p_T$ spectrum for jets from dijet production falls faster than that for jets resulting from 
initial or final state radiation in the $\gamma +$ jets events. Therefore DP fractions depend on 
the jet $p_T$ and the DP fractions as well as $\sigma_{\mbox{\scriptsize eff}}$ can be determined
in data by a set of equations, relating the different contributions in the three second jet transverse 
momentum bins 15 - 20, 20 - 25 and 25 - 30~GeV.
In each momentum bin the discriminating variables
$S_{p_T}=\frac{1}{\sqrt{2}}\sqrt{\left( \frac{|\vec{P_T}(\gamma,i)|}{\delta P_T(\gamma,i)} \right)^2 
+ \left( \frac{|\vec{P_T}(j,k)|}{\delta P_T(j,k)}\right)^2}$,
$S_{p'_T}=\frac{1}{\sqrt{2}}\sqrt{\left( 
 \frac{|\vec{P_T}(\gamma,i)|}{|\vec{P}_T^{\gamma}|+|\vec{P}_T^i|} \right)^2 
 + \left( \frac{|\vec{P_T}(j,k)|}{|\vec{P}_T^j|+|\vec{P}_T^k|}\right)^2}$
and
$S_{\phi}=\frac{1}{\sqrt{2}}\sqrt{\left( \frac{\Delta\phi(\gamma,i)}{\delta \phi(\gamma,i)} \right)^2 
+ \left( \frac{\Delta\phi(j,k)}{\delta \phi(j,k)}\right)^2}$ are evaluated, where $\phi$ is the 
azimuthal angle and $\delta$ indicates the uncertainty of a given variable, defined between pairs 
of photon ($\gamma$) and jet ($i,j,k$) objects.
These variables are computed for the reconstructed particle pairing which minimises $S$. 
For this pairing and each variable the 
final observable $\Delta S = \Delta\phi(\vec{P}_T(\gamma,i),\vec{P}_T(j,k))$ is being determined.

The measurement reveals that the DP fraction drops from 0.47 in the second jet $15<p_T<20$~GeV 
transverse momentum bin to 0.23 in the second jet $25<p_T<30$~GeV transverse momentum bin. Averaging
over the three second jet transverse momentum bins yields 
$<\sigma_{eff}>=16.4\pm 0.3(\mbox{stat})\pm2.3(\mbox{syst})$~mb.
Good agreement with previous CDF measurements in 4-jet~\cite{cdf_4jetDPS_93} and 
$\gamma$+3 jet~\cite{cdf_gammaJet_PDS_97}\cite{cdf_gammaJet_PDS_97_b} events is observed.
From the parameter $\sigma_{\mbox{\scriptsize eff}}$ the proton radius in different parton 
spatial density models (see Table \ref{spatialProton}) can be calculated.

\begin{figure}[t]
\vspace*{-4ex}
\includegraphics[width=7.8cm]{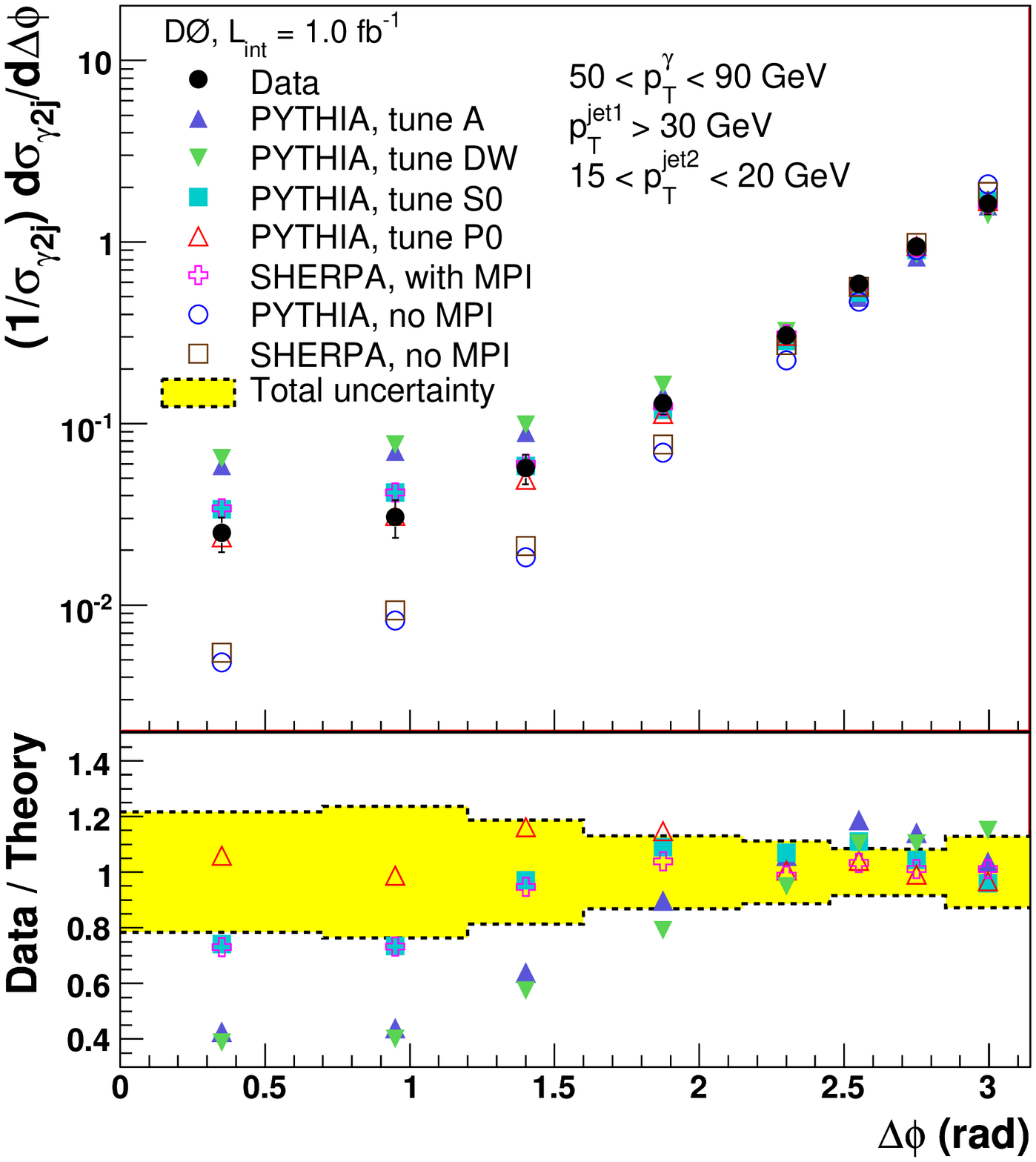}
\includegraphics[width=7.8cm]{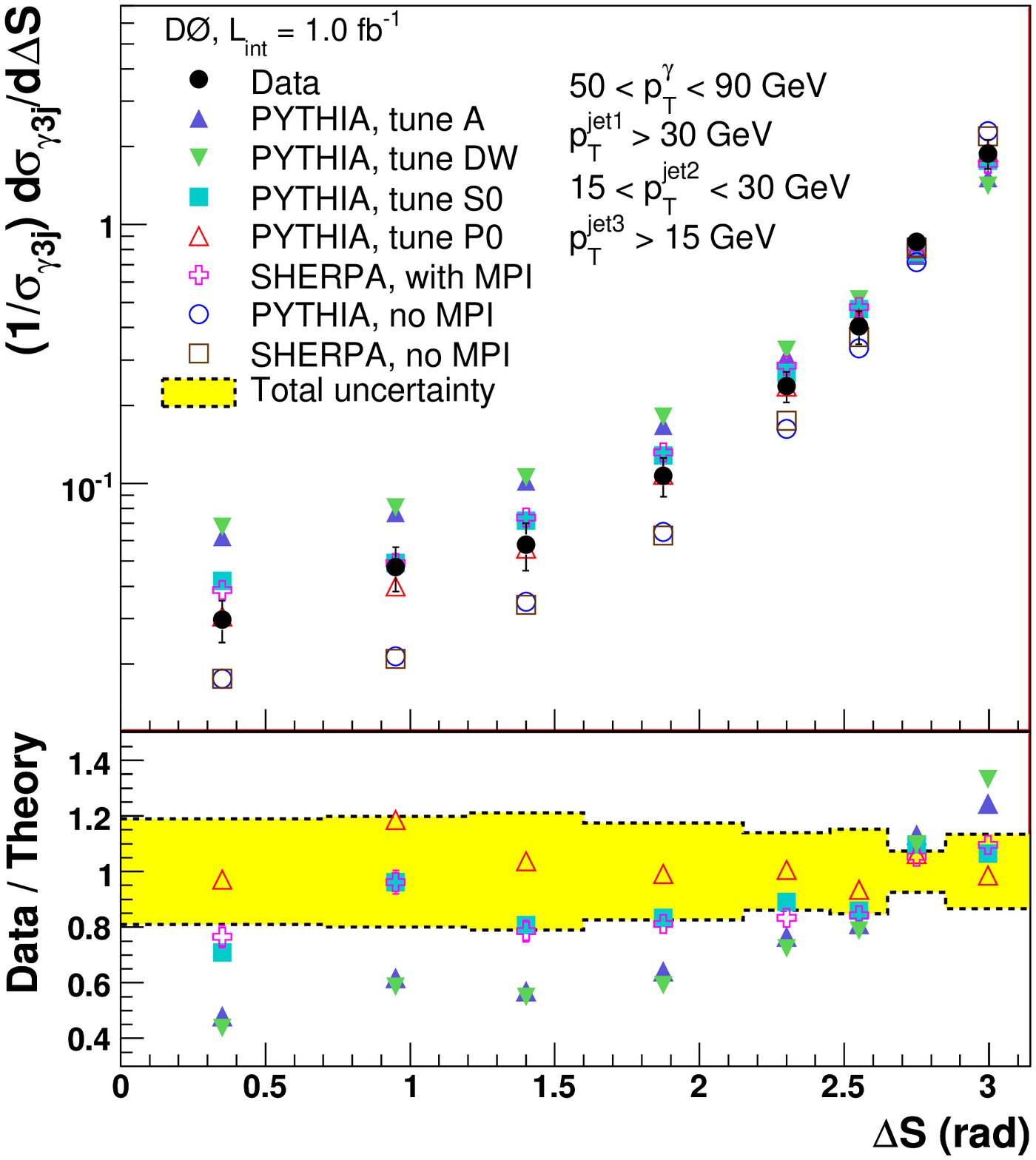}
\vspace*{-3ex}
\caption{ \label{dpsplots}
Differential cross section in $\gamma+2$ jet events as a function of $\Delta \phi$ (left) 
and differential cross section in $\gamma+3$ jet events as a function of $\Delta S$ (right) 
in the second leading jet bin respectively. The data is compared to various models.
}
\end{figure}

\subsection{Azimuthal decorrelations and double parton scattering in $\gamma$ + 2 jet and $\gamma$ + 3 jet events}
The azimuthal decorrelations and double parton scattering in $\gamma$ + 2 jet and $\gamma$ + 3 jet events
are measured~\cite{d0_DPS_gamma2+3jet} based on an integrated luminosity of  1.0 fb$^{-1}$.
This analysis provides an extension of the DPS analysis discussed above in allowing also for
$\gamma$ + 2 jet events, where a third jet is not reconstructed or did not pass the selection criteria.
The gained statistics allows one to subdivide the cross section measurement in bins of the second jet 
transverse momentum and increases thus the sensitivity to MPI models.

Events with an isolated high transverse momentum photon of $50<p_T<90$~GeV are selected. The jet 
transverse momenta are corrected to the hadronic final state. The first jet in transverse momentum 
is required to fulfil $p_T>30$~GeV and one or two more jets with $p_T>15$~GeV have to be found. Jets 
are reconstructed with the iterative midpoint cone jet algorithm~\cite{run2cone} with cone size $R=0.7$.
The event selection for photons and jets is identical to the requirements of the previous analysis 
discussed above. In addition a separation requirement between any pair of selected objects among the
photon and the jets of $\Delta R>0.9$ is used.
The sample of DP candidates is selected from events with a single reconstructed $p\bar{p}$ collision 
vertex. For $\gamma$ + 2 jet events the differential cross section as a function of the azimuthal
angle difference between the photon + leading $p_T$ jet system and the second leading $p_T$ jet
$\Delta\phi(\gamma+\mbox{jet}_1, \mbox{jet}_2)$ is considered in three bins of second leading
$p_T$ jet, namely 15 -20 (see Fig. \ref{dpsplots}, left), 20 - 25 and 25 - 30~GeV.
For $\gamma$ + 3 jet events the differential cross section as a function of the azimuthal
angle difference between the photon  + leading $p_T$ jet system and the second + third leading $p_T$ 
jet $\Delta S = \Delta\phi(\vec{P}_T(\gamma,\mbox{jet}_1),\vec{P}_T(\mbox{jet}_2,\mbox{jet}_3))$,
is measured only in a single 15-30~GeV second transverse momentum jet bin (see Fig. 
\ref{dpsplots}, right) for statistical reasons.
The differential cross sections are normalised for an improved sensitivity to MPI models.
Various models and tunes of PYTHIA~\cite{pythia6420} and SHERPA~\cite{sherpa} are compared to data.
There are large differences between the models and the data confirm the presence of DP events.
The data are close to Perugia (P0)~\cite{P0tune09}, S0~\cite{S0_2005} and SHERPA~\cite{sherpa} 
with MPI tunes, while predictions from previous PYTHIA MPI models with tunes A and DW, making use of
$Q^2$-ordered parton showers, are disfavoured.

\section{Conclusions}
A multitude of D\O\ measurements in the regime of perturbative and non-perturbative QCD has been
elaborated and confronted to theory predictions. While pQCD predictions are in general in agreement 
with the various measurements, limitations of the non-perturbative parts (e.g. PDF's, 
photon fragmentation, hadronisation, underlying event) cannot be hidden and the experimental 
results can be exploited to optimise phenomenological models and their uncertainties.

\vspace*{3ex}


\end{document}